\documentclass[10pt,twoside]{article}
\usepackage{ae} % or {zefonts}
\usepackage[T1]{fontenc}
\usepackage[ansinew]{inputenc}
\usepackage{amsmath}
\usepackage{amssymb}
\usepackage{graphicx}
 \numberwithin{equation}{section}
\begin{document}
%%% ----------------------------------------------------------------------
\title{\large \textbf{HEURISTIC DERIVATION OF BLACKBODY RADIATION LAWS USING
PRINCIPLES OF DIMENSIONAL ANALYSIS}}
\author{\textsc{Gerhard Kramm}\\\\
Geophysical Institute, University of Alaska Fairbanks\\
903 Koyukuk Drive Fairbanks, AK 99775-7320, USA\\\\
\textsc{Fritz Herbert}\\\\
J.W. Goethe-University, Theoretical Meteorology\\
Robert-Mayer-Strasse 1, D-60325 Frankfurt am Main, Germany}
\date{}
%%% ----------------------------------------------------------------------
\maketitle
%%% ----------------------------------------------------------------------
\thispagestyle {headings} \markright{\centerline{To be published in
Journal of the Calcutta Mathematical Society}}
\pagestyle{myheadings} \markboth{\centerline{\textsc{Gerhard Kramm
and Fritz Herbert}}}{\centerline{\textsc{Heuristic Derivation of
Blackbody Radiation Laws Using Principles of\ldots\ldots}}}
%%% ----------------------------------------------------------------------
\begin{abstract}
\noindent A generalized form of Wien's displacement law and the
blackbody radiation laws of (a) Rayleigh and Jeans, (b) Rayleigh,
(c) Wien and Paschen, (d) Thiesen and (e) Planck are derived using
principles of dimensional analysis. This kind of scaling is
expressed in a strictly mathematical manner employing dimensional
$\pi$-invariants analysis sometimes called Buckingham's
$\pi$-theorem. It is shown that in the case of the classical
radiation law of Rayleigh and Jeans only one $\pi$ number occurs
that has to be considered as a non-dimensional universal constant.
This $\pi$ number may be determined theoretically or/and
empirically. It is also shown that dimensional $\pi$-invariants
analysis yields a generalized form of Wien's displacement law. In
this instance two $\pi$ numbers generally occur. Consequently, a
universal function is established that is indispensable to avoid the
so-called \emph{Rayleigh-Jeans catastrophe in the ultraviolet}.
Unfortunately, such a universal function cannot be inferred from
dimensional arguments. It has to be derived theoretically or/and
empirically, too. It is shown that such a similarity function can be
deduced on the basis of heuristic principles, when criteria like the
maximum condition regarding the generalized form of Wien's
displacement law, the requirement of the power law of Stefan and
Boltzmann, and Ehrenfest's arguments regarding the \emph{red} and
the \emph{violet requirements} are adopted.
\end{abstract}
%%% ----------------------------------------------------------------------
%%%                            First Section
%%% ----------------------------------------------------------------------
\setcounter{section}{0} \section{Introduction} \noindent Our
contribution is focused on the heuristic derivation of blackbody
radiation laws, where principles of dimensional analysis are
considered. The idea on which dimensional analysis is based is very
simple. It is inferred from the fact that physical laws do not
depend on arbitrarily chosen basic units of measurements. In
recognizing this simple idea, one may conclude that the functions
that express physical laws must possess a certain fundamental
property, which, from a mathematical point of view, is called the
generalized homogeneity or symmetry \cite{Ba96}. This property
allows the number of arguments in these functions to be reduced,
thereby making it simpler to obtain them. As Barenblatt \cite{Ba96}
pointed out, this is the entire content of dimensional
analysis - there is nothing more to it.\\
\indent Often, solutions for physical problems, especially in
mechanics and fluid mechanics (e.g., \cite{Ba79, Ba94, Ba96, Ba03,
Boh04, Hu52, Ki76, Kram06, Zdu03}) and cloud microphysics
\cite{Bro91}, can be found on the basis of similarity hypotheses
that comprise all problem-relevant dimensional quantities and serve
to possess the physical mechanisms of these problems. Such
similarity hypotheses implicitly describe the functional dependence
between these dimensional quantities in a mathematical form. This
does not mean that this functional dependence can explicitly
substantiated by formulating a similarity hypothesis only. A
similarity hypothesis will become successful if a generalize
homogeneity or a symmetry exist.\\
\indent If similarity is hypothesized, its mathematical treatment
can further be performed by the procedure of dimensional
$\pi$-invariants analysis (sometimes also called the Buckingham's
$\pi$-theorem, for details see \cite{Ba79, Ba94, Ba96, Ba03, Bu14,
Kram06}). During this mathematical treatment the explicit dependence
between the problem-relevant dimensional quantities is expressed by
the non-dimensional $\pi$-invariants.\\
\indent The first attempt to derive a blackbody radiation law on the
basis of dimensional consideration was performed by Jeans
\cite{Je05a, Je06} using the wavelength $\lambda$, the absolute
temperature T, the velocity of light in vacuum c, the charge e and
the mass m of an electron, the universal gas constant R, and the
dielectric constant of the ether K. His attempt to derive Wien's
\cite{Wi94} displacement law, however, was strongly criticized by
Ehrenfest \cite{Eh06a, Eh06b}. Ehrenfest showed that this kind of
dimensional analysis which leads to a similarity function containing
two $\pi$ numbers (see Appendix A) was rather arbitrary. In his
reply Jeans \cite{Je06} rejected Ehrenfest's criticisms. Some months
later, Ehrenfest \cite{Eh06b} also plucked Jeans' additional
arguments to pieces. Since their debate was mainly focus on
dimensional analysis for which a closed theory was not available
during that time (a first step on this road was made by Buckingham
\cite{Bu14}), there was, if at all, only a minor interest on this
debate \cite{Na04}.\\
\indent In the following we will derive a generalized form of Wien's
\cite{Wi94} displacement law and various blackbody radiation laws
from the perspective of dimensional scaling. In doing so, we partly
follow the ideas outlined by Glaser (as cited by \cite{Som56}) and
Sommerfeld \cite{Som56}, i.e., we only postulate that similarity
exists. In the chapter 2, the method used in our dimensional scaling
is presented in a strict mathematical manner. The application of our
method in deriving the blackbody radiation laws is described in
chapter 3. Here, it is shown that in the case of the classical
radiation law of Rayleigh \cite{Ra00, Ra05} and Jeans \cite{Je05b}
only one $\pi$ number occurs that has to be considered as a
non-dimensional universal constant. This $\pi$ number may be
determined theoretically or/and empirically. In the instance of the
generalized form of Wien's \cite{Wi94} displacement law that, in
principle, contains the radiation laws of Wien \cite{Wi96} and
Paschen \cite{Pas96}, Rayleigh \cite{Ra00}, Thiesen \cite{Thi00} and
Planck \cite{Pla00a, Pla00b, Pla01} as special cases, two $\pi$
numbers generally occur. Consequently, a universal function (also
called the similarity function) is established that is indispensable
to avoid the so-called \emph{Rayleigh-Jeans catastrophe in the
ultraviolet} \cite{Eh11}. Unfortunately, such a universal function
cannot explicitly be determined by dimensional $\pi$-invariants
analysis. It has to derive on the basis of theoretical or/and
empirical findings, too. It is shown, however, that such a
similarity function can be deduced on the basis of heuristic
principles, when criteria like the maximum condition regarding the
generalized form of Wien's \cite{Wi94} displacement law, the
requirement of the power law of Stefan \cite{Ste79} and Boltzmann
\cite{Bol84}, and Ehrenfest's \cite{Eh11} \emph{red} and the
\emph{violet requirements} are adopted. Since even Planck's
radiation law can be derived in such a manner, some historical notes
regarding the foundation of the quantum theory are briefly gathered
in chapter 4.
%%% ----------------------------------------------------------------------
%%%                          Second Section
%%% ----------------------------------------------------------------------
\section{Dimensional $\pi$-invariants analysis}
\noindent The theoretical foundation of the procedure is
linked to various sources, for instance, Kitaigorodskij \cite{Ki76},
Barenblatt \cite{Ba79, Ba94, Ba96, Ba03}, Herbert \cite{He80}, Pal
Arya \cite{Pal88}, Brown \cite{Bro91}, Sorbjan \cite{Sor89}, and
Kramm and Herbert \cite{Kram06} which are devoted to characteristic
scaling problems in fluid dynamics and turbulence, boundary layer
meteorology and other physical disciplines. The description mainly
follows the guideline
of Kramm and Herbert \cite{Kram06}.\\
\indent Let adopt that, associated with a certain physical problem,
we can select a set of characteristic dimensionality quantities, for
instance, $\kappa$ variables, parameters or/and constants, $Q_1,
Q_2, \dots, Q_\kappa$ that unambiguously and evidently represent the
arguments of a mathematical relationship. First this "law" is
unspecified; therefore it is formally employed as a general
postulate, commonly referred to as the similarity hypothesis of the
problem, which may read\\
%%% ----------------------------------------------------------------------
\begin{equation}\label{2.1}
F(Q_1, Q_2, \dots, Q_\kappa) = 0\quad.
\end{equation}
%%% ----------------------------------------------------------------------
\noindent In its implicit representation Eq. (\ref{2.1}) declares
$\kappa - 1$ free or independent arguments as well as a
transformation of the full series of $Q_j$ for $j = 1, \dots,
\kappa$ to a series of $p$ non-dimensional invariants $\pi_i$ for $i
= 1, \dots, p$ in terms of a factorization by powers.
Correspondingly,
in that mind each $\pi_i$-expression is defined by\\
%%% ----------------------------------------------------------------------
\begin{equation}\label{2.2}
\pi_i = Q_1^{x_{1,i}}Q_2^{x_{2,i}}\dots Q_\kappa^{x_{\kappa,i}} =
\prod_{j=1}^\kappa Q_j^{x_{j,i}}\quad \textrm{for}\;i = 1,\dots,
p\quad,
\end{equation}
%%% ----------------------------------------------------------------------
\noindent and it is necessarily linked with the condition of
non-dimensionality\\
%%% ----------------------------------------------------------------------
\begin{equation}\label{2.3}
\textrm{dim}\;\pi_i = 1\quad \textrm{for}\;i = 1,\dots, p\quad,
\end{equation}
%%% ----------------------------------------------------------------------
\noindent where $p < \kappa$ is customarily valid.\\
\indent Next, we will suppose that the $\pi$-invariants can have
interdependencies of arbitrary forms, and it may exist a
corresponding relation\\
%%% ----------------------------------------------------------------------
\begin{equation}\label{2.4}
\phi(\pi_1, \pi_2,\dots,\pi_p) = 0
\end{equation}
%%% ----------------------------------------------------------------------
\noindent which is to be understood as an alternative similarity
hypothesis to Eq. (\ref{2.1}). In this function the powers $x_{j,i}$
are basically unknown numbers, and their determination is the proper
problem of the so-called Buckingham $\pi$-theorem. If there are more
than one $\pi$-invariant, i.e., $p > 1$, then we have with Eq.
(\ref{2.4}) the explicit representation\\
%%% ----------------------------------------------------------------------
\begin{equation}\label{2.5}
\pi_i = \varphi(\pi_1, \pi_2,\dots,\pi_p)
\end{equation}
%%% ----------------------------------------------------------------------
\noindent in which $\varphi$ may be interpreted as a universal
function within the framework of the similarity hypothesis, where,
according to the implicit formulation (\ref{2.4}), $\pi_i$ (for any
arbitrary $i\; \epsilon\; \{1, \dots, p\}$) is not an argument of
that universal function. Note that in the special case of $p = 1$,
we will merely obtain one $\pi$-invariant, that is a non-dimensional
universal constant. This special case is expressed by Eq.
(\ref{2.5}) in the singular form\\
%%% ----------------------------------------------------------------------
\begin{equation}\label{2.6}
\pi = \textrm{const.}
\end{equation}
%%% ----------------------------------------------------------------------
\noindent (or $\varphi = \textrm{const.}$). In view to the
determination of the powers $x_{j,i}$, we will extend our treatment
to the concise set of \emph{fundamental} dimensions, $D_n$ for $n =
1,\dots, r$, such as length $L$, time $T$, mass $M$, temperature
$\Theta$, considering that any quantity's dimension can be analysed
in terms of the independent $D_n$ by
homogeneous power factorization. Let that be expressed as\\
%%% ----------------------------------------------------------------------
\begin{equation}\label{2.7}
\textrm{dim}\;Q_j = D_1^{g_{1,j}}D_2^{g_{2,j}}\dots D_r^{g_{r,j}} =
\prod_{n=1}^r D_n^{g_{n,j}}\quad \textrm{for}\;j = 1,\dots,
\kappa\quad,
\end{equation}
%%% ----------------------------------------------------------------------
\noindent in which the powers $g_{n,j}$ for $n = 1, \dots, r$ and $j
= 1, \dots, \kappa$ are known from the relevant quantities $Q_j$
according to the hypothesized similarity condition. Note that $r
\leq \kappa$ is valid, where $r$ is the highest number of
fundamental dimensions that may occur. In other words: for $\kappa$
quantities $Q_j$ including $r$ fundamental dimensions $D_n$ we
obtain $p = \kappa - r$ independent non-dimensional invariants, so-called
$\pi$ numbers.\\
\indent Now a straight-forward development of the analytical
framework is attained by introducing Eq. (\ref{2.7}) together with
the factorization by powers from Eq. (\ref{2.2}) into the condition
of non-dimensionality (\ref{2.3}). In doing so, we obtain this basic
law as described in the following detailed representation\\
%%% ----------------------------------------------------------------------
\begin{equation}\label{2.8}
\textrm{dim}\;\pi_i = \prod_{j=1}^\kappa \Bigl(\prod_{n=1}^r
D_n^{g_{n,j}}\Bigr)^{x_{j,i}} = 1\quad \textrm{for}\;i = 1,\dots,
p\quad.
\end{equation}
%%% ----------------------------------------------------------------------
\noindent Combining the two factorizations $\prod_j$ and $\prod_n$
in this equation enables to rewrite this set of conditions in the
fully equivalent form\\
%%% ----------------------------------------------------------------------
\begin{equation}\label{2.9}
\textrm{dim}\;\pi_i = \prod_{n=1}^r D_n^{\sum_{j=1}^\kappa
g_{n,j}\;x_{j,i}} = 1\quad \textrm{for}\;i = 1,\dots, p\quad.
\end{equation}
%%% ----------------------------------------------------------------------
\noindent For the following conclusion, the latter is more suitable
than the former. Indeed, we may immediately infer from the
factorizing analysis in dependence on the bases $D_n$ for $ n = 1,
\dots, r$, that the set of condition\\
%%% ----------------------------------------------------------------------
\begin{equation}\label{2.10}
\sum_{j=1}^\kappa g_{n,j}\;x_{j,i} = 0\quad \textrm{for}\;n =
1,\dots, r\;\textrm{and}\;i = 1,\dots, p\quad,
\end{equation}
%%% ----------------------------------------------------------------------
\noindent has to hold since each $D_n$-exponential factor must
satisfy, owing to its mathematical independence, the condition of
non-dimensionality (see Eqs. (\ref{2.3}) and (\ref{2.9})), i.e., to
be equal to unity. In matrix notation, Eq. (\ref{2.10}) may be
expressed by\\
%%% ----------------------------------------------------------------------
\begin{equation}\label{2.11}
\underbrace{\left\{
  \begin{array}{cccc}
    g_{1,1} & g_{1,2} & \dots & g_{1,\kappa} \\
    g_{2,1} & g_{2,2} & \dots & g_{2,\kappa} \\
    \dots & \dots & \dots & \dots \\
    g_{r,1} & g_{r,2} & \dots & g_{r,\kappa} \\
  \end{array}
\right\}}_{\begin{array}{c}
               \textrm{dimensional matrix} \\
               \textbf{G} = \{g_{r,\kappa}\} \\
             \end{array}}
\underbrace{\left\{
  \begin{array}{cccc}
    x_{1,1} & x_{1,2} & \dots & x_{1,p} \\
    x_{2,1} & x_{2,2} & \dots & x_{2,p} \\
    \dots & \dots & \dots & \dots \\
    x_{\kappa,1} & x_{\kappa,2} & \dots & x_{\kappa,p} \\
  \end{array}
\right\}}_{\begin{array}{c}
               \textrm{matrix of powers} \\
               \textbf{A} = \{x_{\kappa,p}\} \\
             \end{array}}
= \{0\}\quad,
\end{equation}
%%% ----------------------------------------------------------------------
\noindent where the notation  $\{0\}$ is an $r \times p$ matrix, and
each column of the matrix of powers, $\textbf{A}$, is forming
so-called solution vectors $\textbf{x}_i$ for the invariants $\pi_i$
for $i = 1, \dots, p$. The set of equations (\ref{2.11}) serves to
determine the powers $x_{j,i}$ for $j = 1, \dots, \kappa$, and $i =
1. \dots, p$. So the homogeneous system of linear equations has, in
accord with Eq. (\ref{2.11}), for each of these $\pi$-invariants
the alternative notation\\
%%% ----------------------------------------------------------------------
\begin{equation}\label{2.12}
\textbf{G}\cdot \textbf{x}_i = \textbf{0}\quad \textrm{or}\quad
\left\{
  \begin{array}{cccc}
    g_{1,1} & g_{1,2} & \dots & g_{1,\kappa} \\
    g_{2,1} & g_{2,2} & \dots & g_{2,\kappa} \\
    \dots & \dots & \dots & \dots \\
    g_{r,1} & g_{r,2} & \dots & g_{r,\kappa} \\
  \end{array}
\right\}\left\{
          \begin{array}{c}
            x_{1,i} \\
            x_{2,i} \\
            \dots \\
            x_{\kappa,i} \\
          \end{array}
        \right\}
 = \{0\}\quad \textrm{for}\;i = 1, \dots, p\quad.
\end{equation}
%%% ----------------------------------------------------------------------
\noindent The rank of the dimensional matrix is equal to the number
of fundamental dimensions, $r$. If the number of dimensional
quantities, $\kappa$, is equal to $r$, we will obtain: $p = 0$. In
this case there is only a trivial solution. In the case of $p > 0$,
the homogeneous system of linear equation (\ref{2.12}) is
indeterminate, i.e., more unknowns than equations, a fact that is
true in all instances presented here. Hence, for each of the $p$
non-dimensional $\pi$ numbers, it is necessary to make a reasonable
choice for $p$ of these unknowns, $x_{\kappa,i}$, to put this set of
equations into a solvable state. After that we obtain for each $\pi$
number an inhomogeneous linear equation system that serves to
determine the remaining $r = \kappa - p$ unknowns. Thus, the
remaining $r \times r$ dimensional matrix $\textbf{G}_0 =
\{g_{r,r}\}$ has the rank $r$, too. It is the largest square
sub-matrix for which the determinant is unequal to zero
($|g_{r,r}| \neq 0$). Thus, we have\\
%%% ----------------------------------------------------------------------
\begin{equation}\label{2.13}
\textbf{G}_0\cdot \textbf{x}_i = \textbf{B}_i \quad \textrm{or}\quad
\left\{
  \begin{array}{cccc}
    g_{1,1} & g_{1,2} & \dots & g_{1,r} \\
    g_{2,1} & g_{2,2} & \dots & g_{2,r} \\
    \dots & \dots & \dots & \dots \\
    g_{r,1} & g_{r,2} & \dots & g_{r,r} \\
  \end{array}
\right\}\left\{
          \begin{array}{c}
            x_{1,i} \\
            x_{2,i} \\
            \dots \\
            x_{r,i} \\
          \end{array}
        \right\}
 = \left\{
          \begin{array}{c}
            B_{1,i} \\
            B_{2,i} \\
            \dots \\
            B_{r,i} \\
          \end{array}
        \right\}\quad \textrm{for}\;i = 1, \dots, p\quad.
\end{equation}
%%% ----------------------------------------------------------------------
\noindent This inhomogeneous system of linear equations can be
solved for $x_{m,i}$ for $m = 1, \dots, r$ by employing Cramer's
rule.
%%% ----------------------------------------------------------------------
%%%                           Third Section
%%% ----------------------------------------------------------------------
\section{Similarity hypotheses for blackbody radiation laws}
\noindent In 1860 Kirchhoff \cite{Ki60}
proposed his famous theorem that for any body at a given temperature
the ratio of emissivity, $e_\nu$, i.e., the intensity of the emitted
radiation at a given frequency $\nu$, and absorptivity, $a_\nu$, of
the radiation of the
same frequency is the same generally expressed by\\
%%% ----------------------------------------------------------------------
\begin{equation}\label{3.1}
\frac {e_\nu}{a_\nu} = J(\nu, T)\quad,
\end{equation}
%%% ----------------------------------------------------------------------
\noindent where he called a body perfectly black when $a_\nu = 1$ so
that $J(\nu, T)$ is the emissive power of such a black body \cite{
Ki60, Pai95}. An example of a perfectly black body is the
‘Hohlraumstrahlung’ that describes the radiation in a cavity bounded
by any emitting and absorbing substances of uniform temperature
which are opaque. The state of the thermal radiation which takes
place in this cavity is entirely independent of the nature and
properties of these substances and only depends on the absolute
temperature, $T$, and the frequency (or the wavelength $\lambda$ or
the angular frequency $\omega = 2\; \pi\; \nu$). For this special
case the radiation is homogeneous, isotropic and unpolarized so that
we have \cite{Pai95}\\
%%% ----------------------------------------------------------------------
\begin{equation}\label{3.2}
J(\nu, T) = \frac{c}{8\; \pi}\; U(\nu, T)\quad.
\end{equation}
%%% ----------------------------------------------------------------------
\noindent Here, $U(\nu, T)$ is called the monochromatic (or
spectral) energy density of the radiation in the cavity. Kirchhoff's
theorem has become one of the most general of radiation theory and
expresses the existence of temperature equilibrium for radiation, as
already pointed out by Wilhelm Wien during his Nobel Lecture given
in 1911. Expressions for the monochromatic energy density called the
blackbody radiation laws were derived by (a) Rayleigh \cite{Ra00,
Ra05} and Jeans \cite{Je05b}, (b) Wien \cite{Wi96} and Paschen
\cite{Pas96}, (c) Thiesen \cite{Thi00}, (d) Rayleigh \cite{Ra00},
and (e) Planck \cite{Pla00a, Pla00b, Pla01}.
%%% ----------------------------------------------------------------------
%%%                          Subsection 3.1
%%% ----------------------------------------------------------------------
\subsection{The Rayleigh-Jeans law}
\noindent The dimensional $\pi$-invariants analysis described in
chapter 2 can be employed to derive all these blackbody radiation
laws. First, we consider the radiation law of Rayleigh \cite{Ra00,
Ra05} and Jeans \cite{Je05b} for which the similarity hypothesis
$F(Q_1, Q_2, Q_3, Q_4, Q_5) = F(U, \nu, T, c, k)$ is postulated.
Here, $Q_1 = U$, $Q_2 = \nu$, $Q_3 = T$, $Q_4 = c$, the velocity of
light in vacuum, and $Q_5 = k$, the Boltzmann constant. Obviously,
the number of dimensional
quantities is $\kappa = 5$. The dimensional matrix is given by\\
%%% ----------------------------------------------------------------------
\begin{equation}\label{3.3}
\textbf{G} = \left \{
               \begin{array}{ccccc}
                 - 1 & 0 & 0 & 1 & 2 \\
                 0 & 0 & 1 & 0 & - 1 \\
                 - 1 & - 1 & 0 & - 1 & - 2 \\
                 1 & 0 & 0 & 0 & 1 \\
               \end{array}
             \right \}
\end{equation}
%%% ----------------------------------------------------------------------
\noindent that can be derived from the table of fundamental dimensions,\\
%%% ----------------------------------------------------------------------
\begin{center}
\begin{tabular}{@{\extracolsep{0.05\textwidth}}l|ccccc}
    & U & $\nu$ & T & c & k \\
  \hline
  \textbf{Length} & - 1 & 0 & 0 & 1 & 2 \\
  \textbf{Temperature} & 0 & 0 & 1 & 0 & - 1 \\
  \textbf{Time} & - 1 & - 1 & 0 & - 1 & - 2 \\
  \textbf{Mass} & 1 & 0 & 0 & 0 & 1 \\
\end{tabular}\\
\end{center}
%%% ----------------------------------------------------------------------
\noindent In accord with Eq. (\ref{2.12}), the homogeneous system of
linear equations is given by\\
%%% ----------------------------------------------------------------------
\begin{equation}\label{3.4}
\left \{
  \begin{array}{ccccc}
    - 1 & 0 & 0 & 1 & 2 \\
    0 & 0 & 1 & 0 & - 1 \\
    - 1 & - 1 & 0 & - 1 & - 2 \\
    1 & 0 & 0 & 0 & 1 \\
  \end{array}
\right \} \left\{
  \begin{array}{c}
    x_{1,1} \\
    x_{2,1} \\
    x_{3,1} \\
    x_{4,1} \\
    x_{5,1} \\
  \end{array}
\right\} = \{0\}
\end{equation}
%%% ----------------------------------------------------------------------
\noindent Obviously, the rank of the dimensional matrix is $r = 4$,
and we have $p = \kappa - r = 1$ non-dimensional $\pi$ number
that can be deduced from\\
%%% ----------------------------------------------------------------------
\begin{equation}\label{3.5}
\left.
  \begin{array}{ccccccc}
    -\; x_{1,1} &   &   & +\; x_{4,1} & +\; 2\; x_{5,1} & = & 0 \\
      &   & x_{3,1} &   & -\; x_{5,1} & = & 0 \\
    -\; x_{1,1} & -\; x_{2,1} &   & -\; x_{4,1} & -\; 2\; x_{5,1} & = & 0 \\
    x_{1,1} &   &   &   &  +\; x_{5,1} & = & 0 \\
  \end{array}
\right\}\quad.
\end{equation}
%%% ----------------------------------------------------------------------
\noindent This equation set results in $x_{2,1} = - 2\;x_{1,1}$,
$x_{3,1} = - x_{1,1}$, $x_{4,1} = 3\;x_{1,1}$, and $x_{5,1} = -
x_{1,1}$. Choosing $x_{1,1} = 1$\footnote{) If we choose $x_{1,1} =
\alpha$, where $\alpha \neq 0$ is an arbitrary real number, we will
lead to another invariant $\pi_1^*$. The relationship between this
$\pi$-invariant and that, occurring in Eq. (\ref{3.6}), is given by
$\pi_1 = \sqrt[\alpha]{\pi_1^*}$. Therefore, for convenience, we may
simply choose: $x_{1,1} = 1$.} yields (see Eq. (\ref{3.2}))\\
%%% ----------------------------------------------------------------------
\begin{equation}\label{3.6}
\pi_1 = \prod_{j=1}^5 Q_j^{x_{j,1}} = U^1\; \nu^{-\: 2}\; T^{-\:
1}\; c^3\; k^{-\: 1}\quad.
\end{equation}
%%% ----------------------------------------------------------------------
\noindent Rearranging provides finally\\
%%% ----------------------------------------------------------------------
\begin{equation}\label{3.7}
U(\nu, T) = \pi_1 \frac{\nu^2}{c^3}\; k\; T\quad,
\end{equation}
%%% ----------------------------------------------------------------------
\noindent where the $\pi$-invariant can be identified as $\pi_1 = 8
\pi$. Thus, we have\\
%%% ----------------------------------------------------------------------
\begin{equation}\label{3.8}
U(\nu, T) = \frac{8\; \pi\; \nu^2}{c^3}\; k\;T\quad.
\end{equation}
%%% ----------------------------------------------------------------------
\noindent This radiation law was first derived by Rayleigh
\cite{Ra00, Ra05} using principles of classical statistics, with a
correction by Jeans \cite{Je05b}. Lorentz \cite{Lo03} derived it in
a somewhat different way. Obviously, it fulfills both Kirchhoff's
\cite{Ki60} findings and the requirements of Wien's \cite{Wi94}
conventional displacement law, $U(\nu,T) \propto \nu^3 f(\nu, T)$.
Equation (\ref{3.8}) is called the classical blackbody radiation law
because, as already pointed out by Sommerfeld \cite{Som56}, it is
restricted to the two constants $c$ and $k$ which are well known in
classical physics. In the case when the frequency is small and the
temperature relatively high, this formula works well (indeed it can
be considered as an asymptotic solution for $\nu \rightarrow 0$), as
experimentally proofed by Lummer and Pringsheim \cite{Lu00} and
especially by Rubens and Kurlbaum \cite{Ru00, Ru01}. But it is
obvious that the law of Rayleigh \cite{Ra00, Ra05} and Jeans
\cite{Je05b} cannot be correct because for $\nu \rightarrow \infty$
the monochromatic energy density, $U(\nu,T)$, would tend to
infinity. Ehrenfest \cite{Eh11} coined it the \emph{Rayleigh-Jeans
catastrophe in the ultraviolet}. Consequently, the integral over
$U(\nu,T)$ for all frequencies
yielding the energy density,\\
%%% ----------------------------------------------------------------------
\begin{equation}\label{3.9}
E(T)=\int_{0}^{\infty}U(\nu, T)\;d\nu = \frac{8\; \pi}{c^3} k\;
T\int_{0}^{\infty}\nu^2\;d\nu = \infty\quad,
\end{equation}
%%% ----------------------------------------------------------------------
\noindent would become divergent \cite{Ei05}, in complete contrast
to Boltzmann's \cite{Bol84} thermodynamic derivation of the $T^4$ law,\\
%%% ----------------------------------------------------------------------
\begin{equation}\label{3.10}
E(T) = a\; T^4
\end{equation}
%%% ----------------------------------------------------------------------
\noindent and Stefan's \cite{Ste79} empirical finding
%%% ----------------------------------------------------------------------
\begin{equation}\label{3.11}
R(T) = \frac{c}{4} E(T) = \frac{c\;a}{4} \; T^4 = \sigma\; T^4
\end{equation}
%%% ----------------------------------------------------------------------
\noindent where  $\sigma = c\;a/4$ is customarily called the Stefan
constant.
%%% ----------------------------------------------------------------------
\subsection{A generalized form of Wien's displacement law}
\noindent Since this classical blackbody radiation law completely
fails in the case of high frequencies, it is necessary to look for
an improved similarity hypothesis that must lead to more than one
$\pi$-invariant so that a universal function (or similarity
function) is established. Such a universal function is indispensable
to ensure that the integral in Eq. (\ref{3.9}) stays finite.
However, we cannot introduce another independent variable to get
more than one $\pi$-invariant because it would be in disagreement
with Kirchhoff's \cite{Ki60} findings. Since, on the other hand, the
Rayleigh-Jeans law seems to be reasonable for small frequencies, one
may state that the two dimensional constants of the classical
physics, $c$ and $k$, should also occur in an improved similarity
hypothesis. Consequently, we must look for another dimensional
constant \cite{Som56} and have to make a choice for its fundamental
dimensions. Thus, an improved similarity hypothesis reads
$F(Q_1,Q_2,Q_3,Q_4,Q_5,Q_6) = F(U, \nu, T, c, k, \eta) = 0$, i.e.,
instead of the restriction to the two classical constants $c$ and
$k$, we now have three dimensional constants in our similarity
hypothesis, where the dimensions of the third constant, $Q_6 =
\eta$, is to be chosen in such a sense that $\eta\;\nu/(k\;T)$
becomes non-dimensional, in hoping to establish another
$\pi$-invariant, and, hence, a universal function
$\varphi_R(\eta\;\nu/(k\;T))$, according to the explicit
representation  $\pi_1 = \varphi_R(\pi_2)$ expressed by Eq.
(\ref{2.5}). This means that not only $k\;T$, but also $\eta\;\nu$
have the dimensions of energy. Obviously, the number of dimensional
quantities is $\kappa = 6$, and the table of fundamental dimensions
reads\\
%%% ----------------------------------------------------------------------
\begin{center}
\begin{tabular}{@{\extracolsep{0.05\textwidth}}l|cccccc}
    & U & $\nu$ & T & c & k & $\eta$ \\
  \hline
  \textbf{Length} & - 1 & 0 & 0 & 1 & 2 & 2 \\
  \textbf{Temperature} & 0 & 0 & 1 & 0 & - 1 & 0 \\
  \textbf{Time} & - 1 & - 1 & 0 & - 1 & - 2 & - 1 \\
  \textbf{Mass} & 1 & 0 & 0 & 0 & 1 & 1 \\
\end{tabular}\\
\end{center}
%%% ----------------------------------------------------------------------
\noindent leading to the dimensional matrix:
%%% ----------------------------------------------------------------------
\begin{equation}\label{3.12}
\textbf{G} = \left \{
               \begin{array}{cccccc}
                 - 1 & 0 & 0 & 1 & 2 & 2 \\
                 0 & 0 & 1 & 0 & - 1 & 0 \\
                 - 1 & - 1 & 0 & - 1 & - 2 & - 1 \\
                 1 & 0 & 0 & 0 & 1 & 1 \\
               \end{array}
             \right \}\quad.
\end{equation}
%%% ----------------------------------------------------------------------
\noindent The homogeneous system of linear equations can then be
expressed by (see Eq. (\ref{2.12}))
%%% ----------------------------------------------------------------------
\begin{equation}\label{3.13}
\left \{
  \begin{array}{cccccc}
    - 1 & 0 & 0 & 1 & 2 & 2 \\
    0 & 0 & 1 & 0 & - 1 & 0 \\
    - 1 & - 1 & 0 & - 1 & - 2 & - 1 \\
    1 & 0 & 0 & 0 & 1 & 1 \\
  \end{array}
\right \} \left\{
  \begin{array}{c}
    x_{1,i} \\
    x_{2,i} \\
    x_{3,i} \\
    x_{4,i} \\
    x_{5,i} \\
    x_{6,i} \\
  \end{array}
\right\} = \{0\}\quad \textrm{for}\; i = 1, 2
\end{equation}
%%% ----------------------------------------------------------------------
\noindent where the rank of the dimensional matrix is $r = 4$, i.e.,
we have $p = \kappa - r = 2$  non-dimensional  $\pi$-invariants so
that a universal function, as urgently required, is established.
These two $\pi$ numbers can be derived from\\
%%% ----------------------------------------------------------------------
\begin{equation}\label{3.14}
\left.
  \begin{array}{cccccccc}
    -\; x_{1,i} &   &   & +\; x_{4,i} & +\; 2\; x_{5,i} & +\; 2\; x_{6,i} & = & 0 \\
      &   & x_{3,i} &   & -\; x_{5,i} &   & = & 0 \\
    -\; x_{1,i} & -\; x_{2,i} &   & -\; x_{4,i} & -\; 2\; x_{5,i} & -\; x_{6,i} & = & 0 \\
    x_{1,i} &   &   &   &  +\; x_{5,i} & +\; x_{6,i} & = & 0 \\
  \end{array}
\right\}\quad \textrm{for}\; i = 1, 2\quad.
\end{equation}
%%% ----------------------------------------------------------------------
\noindent Choosing $x_{1,1} = 1$, $x_{6,1} = N$, $x_{1,2} = 0$, and
$x_{6,2} = 1$ yields $x_{2,1} = -2 + N$, $x_{3,1} = -1-N$, $x_{4,1}
= 3$, $x_{5,1} = -1-N$, $x_{2,2} = 1$, $x_{3,2} = -1$, $x_{4,2} =
0$, and $x_{5,2} = -1$. Note that the choice $x_{1,1} = 1$ and
$x_{1,2} = 0$ is indispensable to ensure that $U(\nu,T)$ only occurs
explicitly. Furthermore, $N < 3$ is a real number, and $x_{6,2} = 1$
is chosen in such a sense that $\eta\;\nu/(k\;T)$ is established as
a $\pi$-invariant. According to Eq. (\ref{2.2}), the
$\pi$-invariants are then given by\\
%%% ----------------------------------------------------------------------
\begin{equation}\label{3.15}
\pi_1 = \prod_{j=1}^6 Q_j^{x_{j,1}} = U^1\; \nu^{-\: 2\: +\: N}\;
T^{-\: 1\: -\: N}\; c^3\; k^{-\: 1\; -\: N}\;\eta^N =
\frac{U\;c^3}{\nu^2\;k\;T}\Bigl(\frac{\eta\;\nu}{k\;T}\Bigr)^N
\end{equation}
%%% ----------------------------------------------------------------------
\noindent and\\
%%% ----------------------------------------------------------------------
\begin{equation}\label{3.16}
\pi_2 = \prod_{j=1}^6 Q_j^{x_{j,2}} = U^0\; \nu^1\; T^{-\: 1}\;
c^0\; k^{-\: 1}\;\eta^1 = \frac{\eta\;\nu}{k\;T}\quad.
\end{equation}
%%% ----------------------------------------------------------------------
\noindent According to Eq. (\ref{2.5}), we have\\
%%% ----------------------------------------------------------------------
\begin{equation}\label{3.17}
\pi_1 =
\frac{U\;c^3}{\nu^2\;k\;T}\Bigl(\frac{\eta\;\nu}{k\;T}\Bigr)^N =
\varphi_R(\pi_2) = \varphi_R\Bigl(\frac{\eta\;\nu}{k\;T}\Bigr)\quad,
\end{equation}
%%% ----------------------------------------------------------------------
\noindent where $\varphi_R(\eta\;\nu/(k\;T))$ is the universal
function. Rearranging yields then\\
%%% ----------------------------------------------------------------------
\begin{equation}\label{3.18}
U(\nu,T) =
\frac{\nu^2}{c^3}\;k\;T\;\Bigl(\frac{\eta\;\nu}{k\;T}\Bigr)^{-N}
\varphi_R\Bigl(\frac{\eta\;\nu}{k\;T}\Bigr)\quad.
\end{equation}
%%% ----------------------------------------------------------------------
\noindent For historical reasons and convenience, we may introduce
$8\;\pi$ into Eq. (\ref{3.18}). In doing so, we have\\
%%% ----------------------------------------------------------------------
\begin{equation}\label{3.19}
U(\nu,T) =
\frac{8\;\pi\;\nu^2}{c^3}\;k\;T\;\Bigl(\frac{\eta\;\nu}{k\;T}\Bigr)^{-N}
\Phi_R\Bigl(\frac{\eta\;\nu}{k\;T}\Bigr)\quad,
\end{equation}
%%% ----------------------------------------------------------------------
\noindent where\\
%%% ----------------------------------------------------------------------
\begin{equation}\label{3.20}
\Phi_R\Bigl(\frac{\eta\;\nu}{k\;T}\Bigr) =
\frac{1}{8\;\pi}\;\varphi_R\Bigl(\frac{\eta\;\nu}{k\;T}\Bigr)\quad.
\end{equation}
%%% ----------------------------------------------------------------------
\noindent The merit of Eq. (\ref{3.18}) is that the monochromatic
energy density $U(\nu,T)$ is led to the universal function
$\varphi_R(\eta\;\nu/(k\;T))$ which only depends on the
non-dimensional argument $\pi_2 = \eta\;\nu/(k\;T)$. Obviously, this
equation substantially agrees with Kirchhoff's \cite{Ki60} findings.
Equation (\ref{3.19}) may also be considered as a generalized form
of Wien's \cite{Wi94} displacement law. The conventional form of
this law can be derived by setting $N = - 1$. In doing so, one obtains\\
%%% ----------------------------------------------------------------------
\begin{equation}\label{3.21}
U(\nu,T) = \frac{8\;\pi\;\eta\;\nu^3}{c^3}\;
\Phi_R\Bigl(\frac{\eta\;\nu}{k\;T}\Bigr)\quad.
\end{equation}
%%% ----------------------------------------------------------------------
\subsubsection{Criteria for determining the universal
function heuristically} \noindent As mentioned before, the universal
function $\Phi_R(\eta\;\nu/(k\;T))$ cannot explicitly be inferred
from dimensional arguments. However, any form of such a universal
function must be compatible with following criteria: (a) There must
exist a distinguish maximum. (b)It must guarantee that the integral in\\
%%% ----------------------------------------------------------------------
\begin{equation}\label{3.22}
E(T)= 8\; \pi\;k\;
\Bigl(\frac{k}{c\;\eta}\Bigr)^3\;T^4\int_{0}^{\infty}X^{2\:-\:N}
\Phi_R(X)\;dX
\end{equation}
%%% ----------------------------------------------------------------------
\noindent with $X = \pi_2 = \eta\;\nu/(k\;T)$ keeps finite.
According to this equation, the constant $a$ in Eq. (\ref{3.10}) can
be identified as\\
%%% ----------------------------------------------------------------------
\begin{equation}\label{3.23}
a = 8\; \pi\;k\;
\Bigl(\frac{k}{c\;\eta}\Bigr)^3\int_{0}^{\infty}X^{2\:-\:N}
\Phi_R(X)\;dX\quad.
\end{equation}
%%% ----------------------------------------------------------------------
\noindent Note that Wien \cite{Wi96} also recognized the power law
of Stefan \cite{Ste79} and Boltzmann \cite{Bol84} in deriving the
quantity $c_1\;\nu^3$ that occurs in his radiation law. First, it
must fulfill the requirement that the integral in Eq. (\ref{3.23})
is convergent. If we assume, for instance, $\Phi_R(X) =
X^{N\:-\:3}$, then we will obtain for this integral\\
%%% ----------------------------------------------------------------------
\begin{equation}\label{3.24}
\int_{0}^{\infty}X^{2\:-\:N} \Phi_R(X)\;dX = \mathop{\lim_{b
\:\rightarrow\:\infty}}_{a\: \rightarrow\: 0}
\int_{a}^{b}\frac{dX}{X} = \mathop{\lim_{b
\:\rightarrow\:\infty}}_{a\: \rightarrow\: 0} \ln{\;\frac{b}{a}}
\quad.
\end{equation}
%%% ----------------------------------------------------------------------
\noindent Obviously, this integral is divergent, too. Consequently,
the tendency of the integrand $X^{2\:-\:N} \Phi_R(X)$ to zero when
$X$ approaches to infinity must be faster than $X^{- 1}$. Ehrenfest
\cite{Eh11}, who only considered the case $N = - 1$, postulated
that, therefore, the requirement\\
%%% ----------------------------------------------------------------------
\renewcommand{\theequation}{\arabic{section}.\arabic{equation}a}
\begin{equation}\label{3.25a}
\lim_{\nu\:\rightarrow\:\infty}X^4\;\Phi_R(X) = 0
\end{equation}
%%% ----------------------------------------------------------------------
\noindent must be fulfilled. He coined it the \emph{violet
requirement} (also called the \emph{violet condition} \cite{Kle70}).
For any real number of $N < 3$ we may slightly modify it to\\
%%% ----------------------------------------------------------------------
\addtocounter{equation}{-1}
\renewcommand{\theequation}{\arabic{section}.\arabic{equation}b}
\begin{equation}\label{3.25b}
\lim_{\nu\:\rightarrow\:\infty}X^{3\:-\:N}\;\Phi_R(X) = 0\quad.
\end{equation}
%%% ----------------------------------------------------------------------
\noindent A more rigorous requirement, of course, would be
%%% ----------------------------------------------------------------------
\renewcommand{\theequation}{\arabic{section}.\arabic{equation}}
\begin{equation}\label{3.26}
\lim_{\nu\:\rightarrow\:\infty}X^m\;\Phi_R(X) = 0\quad,
\end{equation}
%%% ----------------------------------------------------------------------
\noindent where $m > 3 - N$. This means that for
$X\rightarrow\infty$ the universal function $\Phi_R(X)$ must tend
with a higher intense to zero than $X^{- m}$ for any $m > 3 - N$.
For $N = - 1$, Ehrenfest \cite{Eh11} coined it the
\emph{strengthened violet requirement}. (c) The energy density,
$U(\nu,T)$, must tend to unity when $\nu$ becomes smaller and
smaller because the Rayleigh-Jeans law (\ref{3.8}) is the asymptotic
solution for that case. Following Ehrenfest \cite{Eh11}, this
requirement may be called the \emph{red requirement} (also called
the \emph{red condition} \cite{Kle70}) being expressed by\\
%%% ----------------------------------------------------------------------
\begin{equation}\label{3.27}
\lim_{\nu\:\rightarrow\:0}\Bigl(\frac{\eta\;\nu}{k\;T}\Bigr)^{-\:N}
\;\Phi_R\Bigl(\frac{\eta\;\nu}{k\;T}\Bigr) = 1\quad.
\end{equation}
%%% ----------------------------------------------------------------------
\subsubsection{The Maximum condition}
\noindent The first derivative of Eq. (\ref{3.19}) reads\\
%%% ----------------------------------------------------------------------
\begin{equation}\label{3.28}
U\textrm{'}(\nu,T) = \frac{\alpha}{\beta^N}\;\nu^{1 - N}\{(2 -
N)\;\Phi_R(\beta\;\nu)+
\beta\;\nu\;\Phi_R\textrm{'}(\beta\;\nu)\}\quad,
\end{equation}
%%% ----------------------------------------------------------------------
\noindent with $\alpha = c_1 = 8\;\pi\;\eta/c^3$ and $\beta =
\eta/(k\;T)$. Thus we obtain for the extreme of $U(\nu,T)$ \\
%%% ----------------------------------------------------------------------
\begin{equation}\label{3.29}
U\textrm{'}(\nu_e,T) = 0\quad\leftrightarrow\quad (2 -
N)\;\Phi_R(\beta\;\nu_e)+
\beta\;\nu_e\;\Phi_R\textrm{'}(\beta\;\nu_e) = 0\quad,
\end{equation}
%%% ----------------------------------------------------------------------
\noindent where $\nu_e$ is the frequency of the extreme. Defining $x
= \beta\;\nu_e$ results in\\
%%% ----------------------------------------------------------------------
\begin{equation}\label{3.30}
(2 - N)\;\Phi_R(x) + x\;\Phi_R\textrm{'}(x) = 0\quad.
\end{equation}
%%% ----------------------------------------------------------------------
\noindent The solution of the equation is given by\\
%%% ----------------------------------------------------------------------
\begin{equation}\label{3.31}
\Phi_R(x) = x^{N - 2}\quad.
\end{equation}
%%% ----------------------------------------------------------------------
\noindent For $\nu_e$ and a given temperature $\Phi_R(x)$ provides a
certain value $Z$. Thus, one obtains\\
%%% ----------------------------------------------------------------------
\begin{equation}\label{3.32}
\frac{\nu_e}{T} = \frac{k}{\eta}\;Z^{\:\tfrac{1}{N - 2}} =
\textrm{const.}
\end{equation}
%%% ----------------------------------------------------------------------
\noindent that may reflect Wien's displacement relation
$\nu_{\textrm{max}}/T = \textrm{const.}$ or
$\lambda_{\textrm{max}}\;T = const.$ Here, $\nu_{\textrm{max}}$ is
the frequency, for which $U(\nu,T)$ reaches its maximum, and
$\lambda_{\textrm{max}}$ is the corresponding wavelength. The second
derivative reads\\
%%% ----------------------------------------------------------------------
\begin{equation}\label{3.33}
U"(\nu,T) = \frac{\alpha}{\beta^N}\;\nu^{-\:N}\;\{(1 - N)(2 -
N)\Phi_R(\beta\;\nu) + 2\;(2 - N)\;\beta\;
\nu\;\Phi_R\textrm{'}(\beta\;\nu) + (\beta\;
\nu)^2\;\Phi_R"(\beta\;\nu)\}\quad.
\end{equation}
%%% ----------------------------------------------------------------------
\noindent Considering Eq. (\ref{3.31}) yields then for the extreme\\
%%% ----------------------------------------------------------------------
\begin{equation}\label{3.34}
U"(\nu_e,T) = 0\quad,
\end{equation}
%%% ----------------------------------------------------------------------
\noindent i.e., we have a stationary point of inflection, rather
than a maximum ($U"(\nu_e,T) < 0$) as requested. This is valid for
any finite value of $N$, i.e., power laws as expressed by Eq.
(\ref{3.31}) are excluded due to the maximum condition. Since the
maximum condition is not generally fulfilled, both Ehrenfest's
\cite{Eh11} statement that Wien's displacement law does not impose
any restriction a priori on the form of $\Phi_R(X)$ and Sommerfeld's
\cite{Som56} derivation of Wien's displacement relation are not
entirely accurate.\\
\indent In contrast to Ehrenfest's statement, the condition
$U"(\nu_e,T) < 0$ clearly imposes a restriction on the form of
$\Phi_R(X)$. For $N < 2$, for instance, the exponential function,\\
%%% ----------------------------------------------------------------------
\begin{equation}\label{3.35}
\Phi_R(X) = \exp{(- X)}\quad,
\end{equation}
%%% ----------------------------------------------------------------------
\noindent guarantees that the maximum condition $U"(\nu_e,T) < 0$ is
always fulfilled (see Appendix B).
%%% ----------------------------------------------------------------------
\subsubsection{Ehrenfest's red and violet requirements and the
blackbody radiation laws of Rayleigh, Wien and Paschen, Thiesen as
well as Planck} \noindent Using this exponential function
(\ref{3.35}), the integral in Eq. (\ref{3.23}) becomes\\
%%% ----------------------------------------------------------------------
\begin{equation}\label{3.36}
\int_0^\infty X^{2 - N}\;\exp{(- X)}\;dX = \Gamma(3 - N)\quad.
\end{equation}
%%% ----------------------------------------------------------------------
\noindent Here, Euler's $\Gamma$-function defined by\\
%%% ----------------------------------------------------------------------
\begin{equation}\label{3.37}
\Gamma(\chi) = \int_0^\infty X^{\chi - 1}\;\exp{(-X)}\;dX
\end{equation}
%%% ----------------------------------------------------------------------
\noindent for all real numbers $\chi = 3 - N > 0$  has been applied.
This condition is clearly fulfilled when the restriction of the
maximum condition, $N < 2$, is considered. Obviously, using this
exponential function guarantees that the integral in Eq.
(\ref{3.23}) keeps finite. It also obeys the strengthened violet
requirement. If we accept this exponential function for a moment, we
will obtain\\
%%% ----------------------------------------------------------------------
\begin{equation}\label{3.38}
U(\nu,T) =
\frac{8\;\pi\;\nu^2}{c^3}\;k\;T\;\Bigl(\frac{\eta\;\nu}{k\;T}\Bigr)^{-N}
\exp{\Bigl(-\:\frac{\eta\;\nu}{k\;T}\Bigr)}\quad.
\end{equation}
%%% ----------------------------------------------------------------------
\noindent For $N = 0$, for instance, we obtain Rayleigh's
\cite{Ra00} radiation formula. In this case the integral in Eq.
(\ref{3.36}) amounts to $\Gamma(3) = 2$. Choosing $N = -\:1$ yields
the radiation law of Wien \cite{Wi96} and Paschen \cite{Pas96} with
$\Gamma(4) = 6$. In the case of Thiesen's \cite{Thi00} radiation
law, which can be derived by setting $N = -\:1/2$, we will obtain
$\Gamma(3.5) = 3.3234$. Obviously, Eq. (\ref{3.38}) contains the
radiation laws of (a) Rayleigh, (b) Wien and Paschen, and (c) Thiesen
 as special cases.\\
\indent Apparently, both the Wien-Paschen radiation law and that of
Thiesen obey the strengthened violet requirement and, of course,
fulfill the maximum condition. However, they do not tend to the
classical blackbody radiation law of Rayleigh and Jeans given by Eq.
(\ref{3.8}), i.e., they do not obey the red requirement. If we
namely express the exponential function by a Maclaurin series, we
will obtain for the Wien-Paschen radiation law ($N = -\:1$)\\
%%% ----------------------------------------------------------------------
\begin{equation}\label{3.39}
exp{(X)} = 1 + X + \frac{X^2}{2} + \frac{X^3}{6} + \ldots
\end{equation}
%%% ----------------------------------------------------------------------
\noindent For small values of $X$ or $\eta\;\nu \ll k\;T$ this
series can be approximated by\\
%%% ----------------------------------------------------------------------
\begin{equation}\label{3.40}
exp{(X)} \cong 1 + X\quad.
\end{equation}
%%% ----------------------------------------------------------------------
\noindent Thus, Eq. (\ref{3.38}) may read\\
%%% ----------------------------------------------------------------------
\begin{equation}\label{3.41}
U(\nu,T) = \frac{8\;\pi\;\nu^2}{c^3}\;k\;T\;\frac{X}{1 + X}\quad.
\end{equation}
%%% ----------------------------------------------------------------------
%%%                             Figure 1
%%% ----------------------------------------------------------------------
\begin{figure}[t]
\begin{center}
\includegraphics[width=.8\textwidth,height=!]{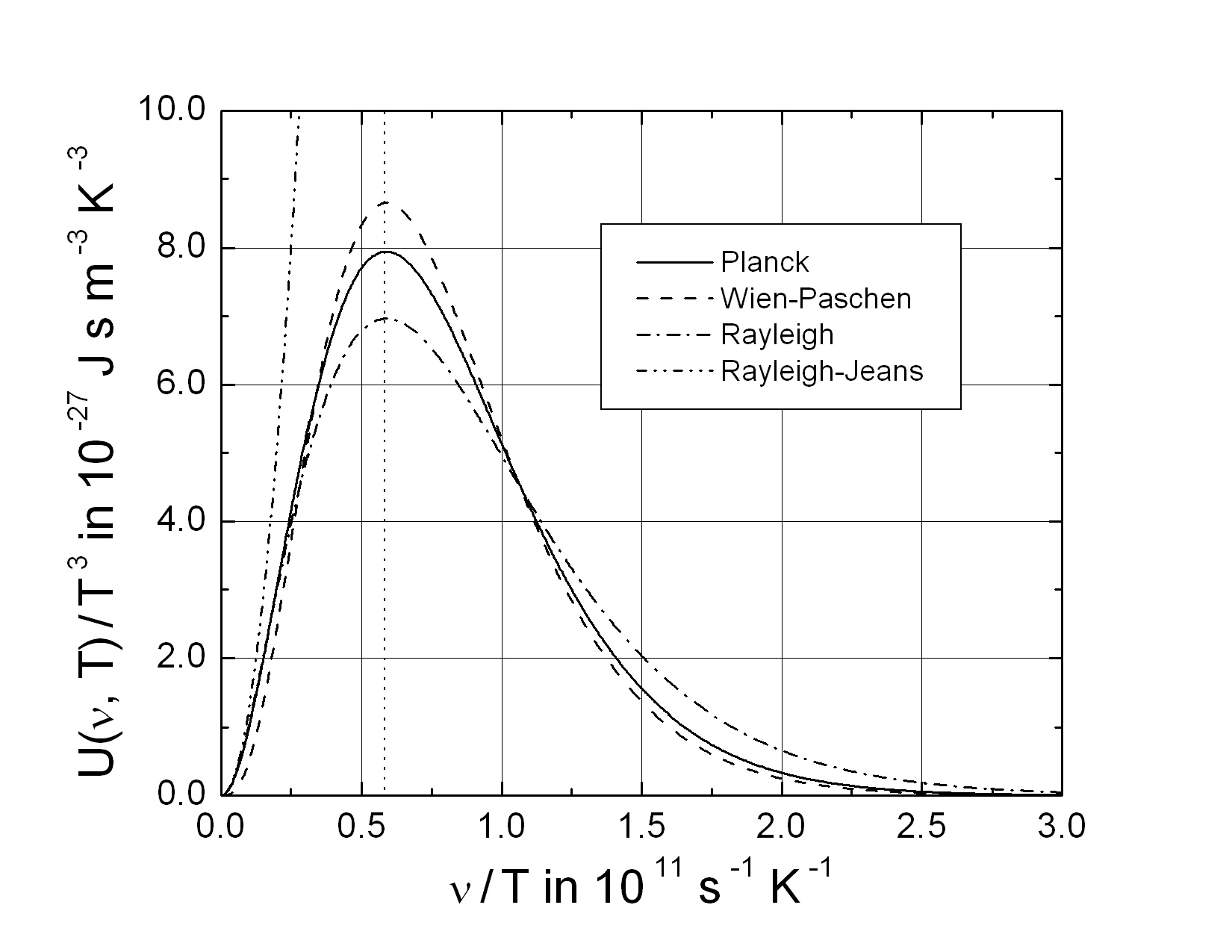}
\end{center}
\caption{Planck's \cite{Pla00a, Pla00b, Pla01} function of the
spectral energy density. The curve represents $U(\nu,T)/T^3$ versus
$\nu/T$ so that it becomes independent of $T$. Note that the
constant, $C = 5.8787\cdot 10^{10}\; s^{-1}\; K^{-1}$, occurring in
Wien's displacement relationship $C = \nu_\textrm{max}/T$ can be
inferred from this figure (dotted line). Also shown are the
functions of Wien \cite{Wi96}, Paschen \cite{Pas96}, Rayleigh
\cite{Ra00}, as well as Rayleigh \cite{Ra00, Ra05} and Jeans
\cite{Je05b}. The constants used for plotting these functions are
listed in Table \ref{Table 1}.} \label{Figure_1}
\end{figure}
%%% ----------------------------------------------------------------------
\noindent Obviously, the expression $X/(1 + X)$ does not converge to
unity, as requested by Eq. (\ref{3.8}) (see also Figure
\ref{Figure_1}). The same is true in the case of Thiesen's radiation
law that can similarly be approximated for that range by\\
%%% ----------------------------------------------------------------------
\begin{equation}\label{3.42}
U(\nu,T) =
\frac{8\;\pi\;\nu^2}{c^3}\;k\;T\;\frac{X^{\tfrac{1}{2}}}{1 +
X}\quad.
\end{equation}
%%% ----------------------------------------------------------------------
\noindent On the contrary, if we choose\\
%%% ----------------------------------------------------------------------
%%%                           Table 1
%%% ----------------------------------------------------------------------
\begin{table}[t]
\caption{Quantities used for plotting the functions illustrated in
Figures \ref{Figure_1} and \ref{Figure_2}.}\label{Table 1}
\begin{center}
\renewcommand\arraystretch{1.8}
\noindent\[
\begin{array}{|l|c|c|c|}
  \hline
         & k & \eta & C \\
  Author & (J\; K^{-1}) & (J\; s) & (s^{-1}\; K^{-1}) \\
  \hline
  Planck & 1.3806 \cdot 10^{-23} & 6.6262 \cdot 10^{-34} & 5.8787 \cdot 10^{10} \\
  \hline
  Wien-Paschen & 1.7963 \cdot 10^{-23} & 9.1670 \cdot 10^{-34} & same \\
  \hline
  Thiesen & 1.8768 \cdot 10^{-23} & 7.9813 \cdot 10^{-34} & same \\
  \hline
  Rayleigh & 1.5967 \cdot 10^{-23} & 5.4323 \cdot 10^{-34} & same \\
  \hline
  Rayleigh-Jeans & 1.3806 \cdot 10^{-23} & 6.6262 \cdot 10^{-34} &   \\
  \hline
\end{array}
\]
\end{center}
\end{table}
%%% ----------------------------------------------------------------------
\begin{equation}\label{3.43}
\Phi_R(X) = \frac{1}{\exp{(X)} - 1}\quad,
\end{equation}
%%% ----------------------------------------------------------------------
\noindent and, again, $N = - 1$, Eq. (\ref{3.19}) will provide\\
%%% ----------------------------------------------------------------------
\begin{equation}\label{3.44}
U(\nu,T) = \frac{8\;\pi\;\eta\;\nu^3}{c^3}\;\frac{1}
{\exp{\Bigl(\dfrac{\eta\;\nu}{k\;T}\Bigr)}- 1}\quad.
\end{equation}
%%% ----------------------------------------------------------------------
%%%                             Figure 2
%%% ----------------------------------------------------------------------
\begin{figure}[t]
\begin{center}
\includegraphics[width=.8\textwidth,height=!]{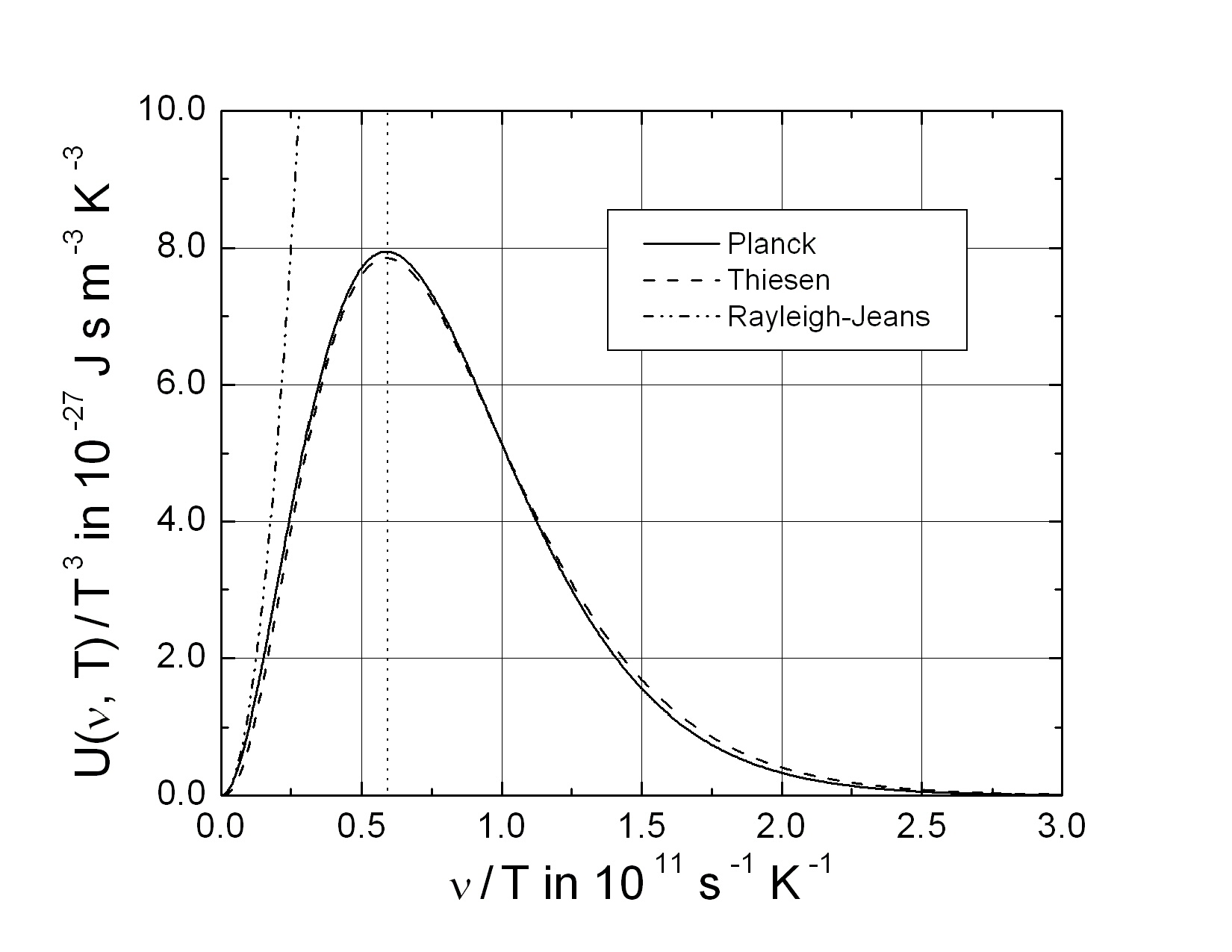}
\end{center}
\caption{As in Figure \ref{Figure_1}, but the functions of Wien
\cite{Wi96}, Paschen \cite{Pas96}, Rayleigh \cite{Ra00} are replaced
by that of Thiesen \cite{Thi00}.} \label{Figure_2}
\end{figure}
%%% ----------------------------------------------------------------------
\noindent This equation is quite similar to Planck's \cite{Pla00a,
Pla00b, Pla01} radiation law. If we again express the exponential
function by a Maclaurin series and again assume $\eta\;\nu \ll
k\;T$, we will obtain\\
%%% ----------------------------------------------------------------------
\begin{equation}\label{3.45}
U(\nu,T) = \frac{8\;\pi\;\nu^2}{c^3}\;k\;T\;\frac{X}{1 + X - 1}
 = \frac{8\;\pi\;\nu^2}{c^3}\;k\;T\quad.
\end{equation}
%%% ----------------------------------------------------------------------
\noindent Apparently, Eq. (\ref{3.44}) completely fulfils the red
requirement (see Figure \ref{Figure_1}). For $\eta\;\nu \gg k\;T$,
i.e., $\exp{(X)} \gg 1$, one also obtains the Wien-Paschen radiation
law. This fact simply states that Eq. (\ref{3.44}) also fulfils the
strengthened violet requirement (see Figure \ref{Figure_1}).\\
\indent Now, we have to check whether the integral in Eq.
(\ref{3.23}) keeps finite. Applying $\Phi_R(X) = (\exp{(X)} - 1)^{-
1}$ and assuming $N = - 1$ yield then\\
%%% ----------------------------------------------------------------------
\begin{eqnarray}\label{3.46}
\nonumber \int_0^\infty \dfrac{X^3}{\exp{(X)} - 1}\; dX  &=&
\int_0^\infty
\dfrac{X^3\exp{(-X)}}{1 - \exp{(-X)}}\; dX \\
\nonumber &=& \int_0^\infty X^3
\exp{(-X)}(1 + \exp{(- X)} + \exp{(-\: 2\:X)} + \ldots)\; dX \\
&=& \int_0^\infty X^3 (\exp{(- X)} + \exp{(-\: 2\:X)} + \exp{(-\:
3\:X)} + \ldots)\; dX \Bigr\}\quad.
\end{eqnarray}
%%% ----------------------------------------------------------------------
\noindent Substituting $n X$ ($n = 1, 2, 3, \ldots$) by $Y$ yields
then\\
%%% ----------------------------------------------------------------------
\begin{equation}\label{3.47}
\int_0^\infty \dfrac{X^3}{\exp{(X)} - 1}\; dX = \sum_{n=1}^\infty
\frac{1}{n^4} \int_0^\infty Y^3 \exp{(- Y)}\;dY = \Gamma(4)
\sum_{n=1}^\infty \frac{1}{n^4} = \frac{\pi^4}{15}\quad,
\end{equation}
%%% ----------------------------------------------------------------------
\noindent where the sum in this equation\\
%%% ----------------------------------------------------------------------
\begin{equation}\label{3.48}
\sum_{n=1}^\infty \frac{1}{n^{2\:k}} = \frac{2^{2\:k\: - 1}\;
\pi^{2\:k}}{(2\;k)!}\;B_k\quad,
\end{equation}
%%% ----------------------------------------------------------------------
\noindent can be calculated using the Bernoulli number for $k = 2$,
namely $B_2 = 1/30$. Apparently, the integral (\ref{3.47}) is
convergent. As $\sigma$ and, hence, $a = 4\;\sigma/c$ are known, the
constant $\eta$ can be determined. In doing so, one obtains: $\eta
=6.6262\cdot 10^{-34}\; J\;s$, i.e., it equals the Planck constant
$h$. Note that Planck called it the "Wirkungsquantum" that means an
elementary quantum of action, and the product $h\;\nu$ is
customarily designated as the \emph{quantum of energy}.
Furthermore, the expression\\
%%% ----------------------------------------------------------------------
\begin{equation}\label{3.49}
\frac{1}{\exp{\Bigl(\dfrac{h\;\nu}{k\;T}\Bigr)}- 1} =
\frac{1}{\exp{\Bigl(\dfrac{\hbar\;\omega}{k\;T}\Bigr)}- 1}
\end{equation}
%%% ----------------------------------------------------------------------
\noindent that occurs in Eq. (\ref{3.44}) is customarily called the
Planck distribution. It may be considered as a special case of the
Bose-Einstein-distribution when the chemical potential of a "gas" of
photons is considered as $\mu = 0$ \cite{Bos24, Ei24, La75}.\\
\indent It is obvious that Planck's \cite{Pla00a, Pla00b, Pla01} law
obeys the requirements in the red range and in the violet range,
and, in addition, it fulfils the maximum conditions. The same, of
course, is also true in the case of Rayleigh's \cite{Ra00} radiation
formula; but as illustrated in Figure \ref{Figure_1} there are
appreciable differences between both radiation laws in the
higher-frequency range. These differences can also be inferred to
the value of the integral in Eq. (\ref{3.36}); for $N = 0$, it
amounts to $\Gamma(3) = 2$. In that range also notable differences
exist between Planck's law and those of Wien \cite{Wi96} and Paschen
\cite{Pas96} because of similar reasons (see Figure \ref{Figure_1}).
Since $\pi^4/15 \cong 6.49$, Planck's distribution yields a value of
the integral in Eq. (\ref{3.23}) which is notably larger than that
provided by this equation when for the asymptotic solution for
$h\;\nu \gg k\;T$, namely the radiation law of Wien and Paschen, $N=
-\:1$, is chosen. As illustrated in Figure \ref{Figure_2}, the
differences between the radiation laws of Planck and Thiesen
\cite{Thi00} are appreciably smaller than in the other cases. These
differences are so small that empirical results alone may be
insufficient for evaluating both radiation laws. Since Thiesen's
radiation law, of course, does not fulfill the red requirement,
Planck's one has to be preferred. Even though the radiation law of
Wien and Paschen represents an asymptotic solution for Planck's law,
the former hardly represents a better solution than Thiesen's one.
One may speculate that the small differences between the radiation
laws of Planck and Thiesen would be ignored today. Note that the
functions of Planck and Rayleigh-Jeans illustrated in Figure
\ref{Figure_1} are based on the values of $k = 1.3806\cdot
10^{-23}\; J\;K^{-1}$ and $\eta = h =6.6262 \cdot 10^{-34} J\;s$
currently recommended. The corresponding values for the functions of
Wien, Thiesen, and Rayleigh have been derived using Stefan's
constant $\sigma = 5.6696 \cdot 10^{-8}\; J m^{-2}\; s^{-1} K^{-4}$
and Wien's relation $\nu_{\textrm{max}}/T = const. = 5.8787 \cdot
10^{10}\; s^{-1}\;K^{-1}$. This constant was calculated using $X =
2.82144$, iteratively computed on the basis of the transcendental
equation $X = 3\;(1 - \exp{(-X)})$ obtained from Planck's radiation
law as the condition for which $U(\nu,T)$ reaches its maximum
\cite{La75}, and the recommend values of $k$ and $h$.
%%% ----------------------------------------------------------------------
%%%                          Fourth Section
%%% ----------------------------------------------------------------------
\section{Some historical notes}
\noindent Planck presented his blackbody radiation law at a meeting
of the German Physical Society on October 19, 1900 in the form of
\cite{Pla00a, La75, Ku78, Pai95, Re95},
%%% ----------------------------------------------------------------------
\begin{equation}\label{4.1}
U(\nu,T) = c_1\;\frac{\nu^3}{\exp{(c_2\;\dfrac{\nu}{T})} - 1}
\end{equation}
%%% ----------------------------------------------------------------------
On the contrary, Eq. (3.44) reflects the form of Planck's radiation
law\footnote{) Note that Planck derived Eq. (\ref{3.44}) using an
equation for the entropy $S$ of a linear harmonic oscillator, and he
related it to its mean energy and to the quantum of energy,
$h\;\nu$.} as presented in his seminal paper published at the
beginning of 1901 \cite{Pla01} and presented to the German Physical
Society on December 14, 1900 \cite{Pla00b, Kle70, Ku78, Pai95,
Re95}. This day may be designated the \emph{birthday of quantum
theory} because the elementary quantum of action explicitly occurred
in Planck's radiation law (e.g., \cite{Ku78, Pai95}. Our use of
principles of dimensional analysis in heuristically deriving Eq.
(\ref{3.44}) by ignoring the aid of the linear harmonic oscillator
model and Planck´s assumption that the energy occurring in
Boltzmann´s \cite{Bol77} distribution is quantized gives evidence
that Planck´s findings were the results of a lucky chance. In his
Nobel Lecture, delivered in 1920, Planck objectively stated: “…even
if the radiation formula should prove itself to be absolutely
accurate, it would still only have, within the significance of a
happily chosen
interpolation formula, a strictly limited value.”\\
\indent The first who, indeed, realized the true nature of Planck's
constant was Einstein \cite{Ei05}. In his article he related the
monochromatic radiation, from a thermodynamic point of view, to
mutually independent light quanta (or photons) and their magnitude,
in principle, to $\Delta\varepsilon = h\;\nu$ that occurs in
Planck's radiation law (see Eq. (\ref{3.44})). Einstein, however,
did not deal with Planck's radiation law in its exact manner, but
rather with the approximation that fits the radiation law of Wien
\cite{Wi96} and Paschen \cite{Pas96}. It seems that he first
recognized that the quantum discontinuity was an essential part of
Planck's radiation theory (e.g., \cite{Da92, Krah99, Ku78}). As
discussed by Klein \cite{Kle70} and Navarro and Pérez \cite{Na04}, a
milestone on the road of quantum discontinuity and light quanta is
Ehrenfest's \cite{Eh11} article on the essential nature of the
different quantum hypotheses in radiation theory. With his article
Ehrenfest contributed to the clarification of the hypothesis of
light quanta. Unfortunately his contribution was not recognized for
a long time.
%%% ----------------------------------------------------------------------
%%%                              Appendix
%%% ----------------------------------------------------------------------
\appendix
\renewcommand{\thesection}{\large Appendix\;\Alph{section}:}
\renewcommand{\theequation}{A\arabic{equation}}
\section{\large Jeans' attempt to derive Wien's
displacement law using dimensional analysis} \noindent Following
Jeans \cite{Je05a, Je06}, the similarity hypothesis reads $F(Q_1,
Q_2, Q_3, Q_4, Q_5, Q_6, Q_7, Q_8) = F(U, \lambda, T, c, e,$ $m, R,
K)= 0$. Here, $Q_2 = \lambda$ is the wavelength, $Q_5 = e$ and $Q_6
= m$ are the charge and the mass of an electron, respectively, $Q_7
= R$ is the universal gas constant, and $Q_8 = K$ the dielectric
constant of the ether expressed with respect to an arbitrary
measuring system. All other symbols have the same meaning as
mentioned before. Since the monochromatic energy density has to be
considered, the number of dimensional quantities
is $\kappa = 8$ now. The dimensional matrix is given by\\
%%% ----------------------------------------------------------------------
\begin{equation}\label{A1}
\textbf{G} = \left \{
  \begin{array}{cccccccc}
    - 1 & 1 & 0 & 1 & 3/2 & 0 & 2 & 0 \\
    0 & 0 & 1 & 0 & 0 & 0 & - 1 & 0 \\
    - 1 & 0 & 0 & - 1 & - 1 & 0 & - 2 & 0 \\
    1 & 0 & 0 & 0 & 1/2 & 1 & 1 & 0 \\
    0 & 0 & 0 & 0 & 1/2 & 0 & 0 & 1 \\
    \end{array}
  \right \}
\end{equation}
%%% ----------------------------------------------------------------------
\noindent that can be inferred from the table of fundamental
dimensions given by\\
%%% ----------------------------------------------------------------------
\begin{center}
\begin{tabular}{@{\extracolsep{0.04\textwidth}}l|cccccccc}
    & U & $\lambda$ & T & c & e & m & R & K \\
  \hline
  \textbf{Length} & - 1 & 1 & 0 & 1 & 3/2 & 0 & 2 & 0 \\
  \textbf{Temperature} & 0 & 0 & 1 & 0 & 0 & 0 & - 1 & 0 \\
  \textbf{Time} & - 1 & 0 & 0 & - 1 & - 1 & 0 & - 2 & 0\\
  \textbf{Mass} & 1 & 0 & 0 & 0 & 1/2 & 1 & 1 & 0 \\
  \textbf{Arbitrary system} & 0 & 0 & 0 & 0 & 1/2 & 0 & 0 & 1 \\
\end{tabular}
\end{center}
%%% ----------------------------------------------------------------------
\noindent The homogeneous system of linear equations can then be
written as (see Eq. (\ref{2.12}))\\
%%% ----------------------------------------------------------------------
\begin{equation}\label{A2}
\left \{
  \begin{array}{cccccccc}
    - 1 & 1 & 0 & 1 & 3/2 & 0 & 2 & 0 \\
    0 & 0 & 1 & 0 & 0 & 0 & - 1 & 0 \\
    - 1 & 0 & 0 & - 1 & - 1 & 0 & - 2 & 0 \\
    1 & 0 & 0 & 0 & 1/2 & 1 & 1 & 0 \\
    0 & 0 & 0 & 0 & 1/2 & 0 & 0 & 1 \\
  \end{array}
  \right \} \left\{
  \begin{array}{c}
    x_{1,i} \\
    x_{2,i} \\
    x_{3,i} \\
    x_{4,i} \\
    x_{5,i} \\
    x_{6,i} \\
    x_{7,i} \\
    x_{8,i} \\
  \end{array}
\right\} = \{0\}\quad \textrm{for}\; i = 1, 2, 3
\end{equation}
%%% ----------------------------------------------------------------------
\noindent Since the rank of the dimensional matrix is $r = 5$, we
have  $p = \kappa - r = 3$ non-dimensional $\pi$-invariants. Thus, a
universal function of the form $\pi_1 = \varphi(\pi_2, \pi_3)$ is
established. These three $\pi$-invariants can be derived from\\
%%% ----------------------------------------------------------------------
\begin{equation}\label{A3}
\left.
  \begin{array}{cccccccccc}
    -\; x_{1,i} & +\; x_{2,i} &   & +\; x_{4,i} & +\; 3/2\; x_{5,i} &  & +\; 2\; x_{7,i} &   & = & 0 \\
      &   & x_{3,i} &   &   &   & -\; x_{7,i} &   & = & 0 \\
    -\; x_{1,i} &   &   &  -\; x_{4,i} & -\; x_{5,i} &   & -\; 2\; x_{7,i} &   & = & 0 \\
    x_{1,i} &   &   &   &  +\; 1/2\; x_{5,i} & +\; x_{6,i} & +\; x_{7,i} &   & = & 0 \\
      &   &   &   &  1/2\; x_{5,i} &   &   & +\; x_{8,i} & = & 0 \\
  \end{array}
\right\}\quad \textrm{for}\; i = 1, 2, 3\quad.
\end{equation}
%%% ----------------------------------------------------------------------
\noindent Choosing $x_{1,1} = 1$, $x_{6,1} = 0$, $x_{7,1} = - 1$,
$x_{1,2} = 0$, $x_{6,2} = - 1$, $x_{7,2} = 1$, $x_{3,1} = 0$,
$x_{6,3} = - 2$ and $x_{7,3} = 1$ yields $x_{2,1} = 2$, $x_{3,1} = -
1$, $x_{4,1} = 1$, $x_{5,1} = 0$, $x_{8,1} = 0$, $x_{2,2} = 0$,
$x_{3,2} = 1$, $x_{4,2} = - 2$, $x_{5,2} = 0$, $x_{8,2} = 0$,
$x_{2,3} = -1$, $x_{3,3} = 1$, $x_{4,3} = -4$, $x_{5,3} = 2$, and
$x_{8,3} = -1$. The choice $x_{1,1} = 1$, $x_{1,2} = 0$, and
$x_{1,3} = 0$ is required to guarantee that $U(\lambda, T)$ only
occurs explicitly. In accord with Eq. (\ref{2.2}), the
$\pi$-invariants are given by\\
%%% ----------------------------------------------------------------------
\begin{equation}\label{A4}
\pi_1 = \prod_{j=1}^8 Q_j^{x_{j,1}} = U^1\; \lambda^2\; T^{-\: 1}\;
c^1\; e^0\; m^0\; R^{-\:1}\; K^0 =
\frac{U\;\lambda^2\;c}{R\;T}\quad,
\end{equation}
%%% ----------------------------------------------------------------------
\begin{equation}\label{A5}
\pi_2 = \prod_{j=1}^8 Q_j^{x_{j,2}} = U^0\; \lambda^0\; T^1\;
c^{-\:2}\; e^0\; m^{-\:1}\; R^1\; K^0 = \frac{R\;T}{m\;c^2}\quad,
\end{equation}
%%% ----------------------------------------------------------------------
\noindent and\\
%%% ----------------------------------------------------------------------
\begin{equation}\label{A6}
\pi_3 = \prod_{j=1}^8 Q_j^{x_{j,3}} = U^0\; \lambda^{-1}\; T^1\;
c^{-4}\; e^2\; m^{-\:2}\; R^1\; K^{-1} = \frac{R\;T\;e^2}{\lambda\;
m^2\;c^4\;K}\quad.
\end{equation}
%%% ----------------------------------------------------------------------
\noindent According to Eq. (\ref{2.5}), we have\\
%%% ----------------------------------------------------------------------
\begin{equation}\label{A7}
\pi_1 = \frac{U\;\lambda^2\;c}{R\;T} =
\varphi\Bigl\{\frac{R\;T}{m\;c^2}, \frac{R\;T\;e^2}{\lambda\;
m^2\;c^4\;K}\Bigr\}
\end{equation}
%%% ----------------------------------------------------------------------
\noindent or\\
%%% ----------------------------------------------------------------------
\begin{equation}\label{A8}
U(\lambda,T) = \frac{R\;T}{\lambda^2\;c}\;
\varphi\Bigl\{\frac{R\;T}{m\;c^2}, \frac{R\;T\;e^2}{\lambda\;
m^2\;c^4\;K}\Bigr\}\quad.
\end{equation}
%%% ----------------------------------------------------------------------
\noindent It is apparent that Eq. (\ref{A8}) completely disagrees
with\\
%%% ----------------------------------------------------------------------
\begin{equation}\label{A9}
U(\lambda,T) \propto \frac{T}{\lambda^4}\; \varphi_R(\lambda,T)
\end{equation}
%%% ----------------------------------------------------------------------
\noindent or\\
%%% ----------------------------------------------------------------------
\begin{equation}\label{A10}
U(\lambda,T) \propto \lambda^{-\:5}\; \varphi_R(\lambda,T)
\end{equation}
%%% ----------------------------------------------------------------------
\noindent that can be derived from Eqs. (\ref{3.19}) and
(\ref{3.21}), respectively. Since we always have $x_{2,i} =
2\;x_{1,i}$, any other choice can give no better relationships.
Therefore, Jeans' attempt to derive Wien's \cite{Wi94} displacement
law completely fails because of his inadequate similarity
hypothesis.
%%% ----------------------------------------------------------------------
\renewcommand{\theequation}{B\arabic{equation}}
\section{\large Derivation of Wien's displacement
relation for $\Phi_R(X)=\exp{(-X)}$} \noindent If we consider the
exponential function $\Phi_R(\beta\;\nu)=\exp{(- \beta\;\nu)}$, the
first two
derivatives of Eq. (\ref{3.19}) will read\\
%%% ----------------------------------------------------------------------
\begin{equation}\label{B1}
U\textrm{'}(\nu,T) = \frac{\alpha}{\beta^N}\;\nu^{1 - N}\;(2 - N -
\beta\;\nu)\;\exp{(-\beta\;\nu)}
\end{equation}
%%% ----------------------------------------------------------------------
\noindent and\\
%%% ----------------------------------------------------------------------
\begin{equation}\label{B2}
U"(\nu,T) = \frac{\alpha}{\beta^N}\;\nu^{-\:N}\;\{(1 - N)(2 - N) -
2\;(2 - N)\;\beta\;\nu + (\beta\;\nu)^2\}\;\exp{(-\beta\;\nu)}\quad.
\end{equation}
%%% ----------------------------------------------------------------------
\noindent For the extreme we obtain $\beta\;\nu_e = 2 - N$ or\\
%%% ----------------------------------------------------------------------
\begin{equation}\label{B3}
\frac{\nu_e}{T} = (2 - N)\;\frac{k}{\eta} = \textrm{const.}
\end{equation}
%%% ----------------------------------------------------------------------
\noindent Introducing Wien's displacement relation into Eq.
(\ref{B2}) yields finally\\
%%% ----------------------------------------------------------------------
\begin{equation}\label{B4}
U"(\nu_e,T) = (N - 2)\;\frac{\alpha}{\beta^N}\;\exp{(N - 2)}\quad,
\end{equation}
%%% ----------------------------------------------------------------------
\noindent This means that $U"(\nu_e,T)$ becomes negative for any
real value of $N$ that fulfils the condition $N < 2$, i.e., in such
case the condition of a maximum is fulfilled.
%%% ----------------------------------------------------------------------
%%%                              Acknoledgement
%%% ----------------------------------------------------------------------
\paragraph{Acknowledgement.} \noindent We would like to express our thanks
to Prof. Dr. habil. Nicole Mölders from the University of Alaska
Fairbanks for her valuable review of the manuscript and fruitful
recommendations.
%%% ----------------------------------------------------------------------
%%%                              References
%%% ----------------------------------------------------------------------

\end{document}